\documentclass{phbauth}
\usepackage{graphicx}
\begin{document}
\begin{frontmatter}
\title{Classical and quantum magnetisation reversal studied in single nanometer-sized 
particles and clusters using micro--SQUIDs}

\author[address1]{W. Wernsdorfer\thanksref{thank1}},
\author[address1]{E. Bonet Orozco},
\author[address1]{B. Barbara}
\author[address2]{A. Benoit}
\author[address3]{D. Mailly}

\address[address1]{Lab. L. N\'eel - CNRS, BP166, 38042 Grenoble, France}
\address[address2]{CRTBT -- CNRS, BP166, 38042 Grenoble, France}
\address[address3]{LMM -- CNRS, 196 av. H. Ravera, 92220 Bagneux, France}

\thanks[thank1]{Corresponding author. E--mail: wernsdor@labs.polycnrs-gre.fr} 

\begin{abstract}
Recent progress in experiment on quantum tunnelling of the magnetic moment 
in mesoscopic systems will be reviewed. The emphasis will be made on measurements of 
individual nanoparticles. These nanomagnets allow one to test the border between classical 
and quantum behaviour. Using the micro--SQUID magnetometer, waiting time, switching 
field and telegraph noise measurements show unambiguously that the magnetisation 
reversal of small enough single crystalline nanoparticles is 
described by a model of thermal activation 
over a single--energy barrier. Results on insulating BaFeO nanoparticles show strong 
deviations from this model below 0.4 K which agree with the theory of macroscopic quantum 
tunnelling in the low dissipation regime.
\end{abstract}

\end{frontmatter}

\section{Introduction}
Macroscopic quantum tunnelling (MQT) represents one of the most fascinating 
phenomena in condensed matter physics. MQT means the tunnelling of a 
macroscopic variable through a barrier characterized by an 
effective potential of a macroscopic system. 
Today, macroscopic systems showing amazing 
agreements with theoretical predictions are 
Josephson Junctions and SQUIDs. Here, quantum tunnelling between two 
macroscopically distinct current states has been observed. After a slow evolution during 
the last ten years, MQT in magnetism now constitutes a new and very interesting field of 
research. It has been predicted that MQT can be observed in magnetic systems with low 
dissipation. In this case, it is the tunnelling of the 
magnetisation vector of a single--domain 
particle through its anisotropy energy barrier or the tunnelling 
of a domain wall through its pinning energy. 
These phenomena have been studied theoretically and experimentally 
\cite{Chichilianne94}.

This brief review focuses on MQT studied in individual 
nanoparticles or nanowires where 
the complications due to distributions of particle size, shape etc. are avoided. The 
experimental evidence of MQT in a single--domain particle or in assemblies of particles is 
still a controversial subject. Therefore, we pay most attention 
to the necessary experimental 
conditions for MQT and review some experimental results. 
We start by reviewing some 
important predictions concerning MQT in a single--domain particle.

\section{Magnetisation reversal by quantum tunnelling}

On the theoretical side, it was shown that in small magnetic particles, a macroscopically 
large number of spins coupled by strong exchange interaction, can tunnel through the energy barrier created by magnetic anisotropy. It has been proposed that there is a characteristic crossover temperature $T_{\rm c}$ below which the 
escape of the magnetisation from a metastable state is dominated by quantum barrier 
transitions, rather than by thermal over barrier activation. 
Above $T_{\rm c}$ the escape rate is 
given by thermal over barrier activation.

In order to compare experiments with theories, predictions of the crossover temperature 
$T_{\rm c}$ and the escape rate $\Gamma_{\rm QT}$ 
in the quantum regime are relevant. Both 
variables should be expressed as a function of parameters which can be changed 
experimentally. Typical parameters are the number of spins $S$, effective anisotropy 
constants, applied field strength and direction, etc.
Many theoretical papers have been published during the last few years 
\cite{Chichilianne94}. We discuss here a result specially adapted for single particle 
measurements which concerns the field dependence 
of the crossover temperature $T_{\rm c}$.

\begin{figure}[b]
\begin{center}\leavevmode
\includegraphics[width=0.8\linewidth]{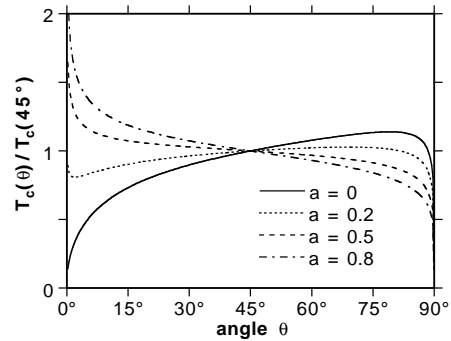}
\caption{Normalised crossover temperature $T_{\rm c}$ as given by equation. (1), and for 
several values of the ratio $a = H_{\perp}/H_{\parallel}$.}
\label{fig1}
\end{center}
\end{figure}

The crossover temperature $T_{\rm c}$ can be 
defined as the temperature where the quantum 
switching rate equals the thermal one. The case of a magnetic particle, as a 
function of the applied field direction, was considered by several authors \cite{Zaslavskii90,Miguel96,Kim97}. We have chosen 
the result for a particle with biaxial anisotropy as the 
effective anisotropy of most particles can be approximately 
described by a strong uniaxial 
and a weak transverse anisotropy. 
The result of Kim can be written in the following form~\cite{Kim97}: 
\begin{eqnarray}
T_{\rm c}(\theta) & \sim & \mu_0 H_{\parallel} \varepsilon^{1/4}\mid cos\theta 
\mid^{1/6}\left(1+
\mid cos\theta \mid^{2/3}\right)^{-1} \nonumber \\
  &    & \mbox{} \qquad\ast\sqrt{1+a\left(1+\mid cos\theta \mid^{2/3}\right)}
\label{eq1}
\end{eqnarray}
where $\mu_0 H_{\parallel}$ and $\mu_0 H_{\perp}$ are the parallel and transverse 
anisotropy field given in Tesla, $\theta$ is the angle between the easy axis of 
magnetisation and the direction of the applied field, and $\varepsilon = (1-
H/H_{\rm sw}^0)$. $H_{\rm sw}^0$ is the 
classical switching field at zero temperature~\cite{St-W48}.
Equation (\ref{eq1}) is valid for any ratio $a = H_{\perp}/H_{\parallel}$. 
The proportionality coefficient of Equation (\ref{eq1}) 
is of the order of unity ($T_{\rm c}$ is in units 
of Kelvin) and depends on the approach of the calculation \cite{Kim97}. Equation (\ref{eq1}) is plotted in Fig.\ \ref{fig1} for several values of the ratio $a$. 
It is valid in the range $\sqrt{\varepsilon} < \theta < \pi/2 - \sqrt{\varepsilon}$. 
%

The most interesting feature which may be drawn from equation (\ref{eq1}) is that the 
crossover temperature is tuneable using the external field strength and direction (Fig.\ 
\ref{fig1}) because the tunnelling probability is increased by the transverse component of 
the applied field. Although at high transverse fields, $T_{\rm c}$ decreases again due to a 
broadening of the anisotropy barrier. Therefore, quantum tunnelling experiments should 
always include studies of angular dependencies. When the effective magnetic anisotropy 
of the particle is known, MQT theories give clear predictions with no fitting parameters.

MQT should also be studied as a function of the effective magnetic anisotropy. In 
practice, it is well known for single particle measurements that each particle 
is in reality somewhat different. Therefore, the effective magnetic 
anisotropy has to be determined for each individual 
particle.

In general, $T_{\rm c}$ does not depend directly on the particle 
volume because $T_{\rm c}$ is 
proportional to the barrier and inversely proportional to the total spin of the particle. 
However, there is a slight volume dependence arising 
form the $\varepsilon-$dependence. 
As experiments are always limited by a certain time window and as the tunnelling rate is 
exponentially dependent on the particle volume, higher applied field values are required for 
larger particles to see the tunnelling in the given experimental time window, i.e. 
$\varepsilon$ is small for larger particles. Moreover, the magnetisation reversal of larger 
particles becomes slightly incoherent, destroying MQT effects. Therefore, the volume 
dependence (i.e. $\varepsilon-$dependence) of MQT should be studied in the particle size 
range where magnetisation reversal by uniform rotation exists.

Finally, it is important to note that most of the MQT theories neglect damping 
mechanisms. We discussed the case of ohmic damping in Ref. \cite{Coffey98} which is the simplest form of damping. More complicated damping mechanisms are not excluded and 
might play an important role. We expect more theoretical work on this in future.

\section{Brief review of magnetisation measurements of individual single--domain 
nanoparticles and wires at very low temperatures}

In order to avoid the complications due to distributions of particle size, shape etc., some 
groups tried to study the thermal and field dependence of magnetisation reversal of 
individual magnetic particles or wires. Most of the recent studies were done by using 
Magnetic Force Microscopy at room temperature. Low temperature investigations were 
mainly performed via resistance measurements.

The first magnetisation measurements of individual single--domain nanoparticles at low 
temperature (0.1 $<$ T(K) $<$ 6) were presented by Wernsdorfer {\it et al.} \cite{Wernsdorfer95}. The detector (a Nb micro-bridge-DC-SQUID) 
and the particles studied (ellipses with axes between 50 and 1000 nm and 
thickness between 5 and 50 nm) were fabricated using electron beam lithography. Electrodeposited wires (with diameters ranging from 40 to 100 nm and lengths up to 5000 nm) 
were also studied \cite{Wernsdorfer96}. Waiting time and switching 
field measurements showed that the magnetisation reversal of these particles and wires 
results from a single thermally activated domain wall nucleation, followed by a fast wall 
propagation reversing the particle's magnetisation. For nanocrystalline Co particles of 
about 50 nm and below 1 K, a flattening of the temperature dependence of the mean 
switching field was observed which could not be explained by thermal activation. These 
results were discussed in the context of MQT of magnetisation. However the width of the 
switching field distribution and the probability of switching are in disagreement with such 
a model because nucleation is very sensitive to factors such as surface defects, surface 
oxidation and perhaps nuclear spins. The fine structure of pre-reversal magnetisation 
states is then governed by a multivalley energy landscape (in a few cases distinct 
magnetisation reversal paths were effectively observed \cite{Wernsdorfer95} and the 
dynamics of reversal occurs via a complex path in configuration space. 

Coppinger {\it et al.} \cite{Coppinger95} used telegraph noise 
spectroscopy to investigate the 
two--level fluctuations (TLF) observed in the conductance of a sample containing 
self-organising ErAs quantum wires and dots in a semi-insulating GaAs matrix. They showed 
that the TLF could be related to two possible magnetic states of a ErAs cluster and that the 
energy difference between the two states was a linear function of the magnetic field. They 
deduced that the ErAs cluster should contain a few tens of Er atoms. At temperatures 
between 0.35~K and 1~K, the associated switching rate of the TLF were thermally activated, 
however below 350 mK the switching rate became temperature independent. Tunnelling of 
the magnetisation was proposed in order to explain the observed behaviour. 

Some open questions remain: What is the object which is really probed by TLF? If this is 
a single ErAs particle, as assumed by the authors, the switching probability  should be an 
exponential function of time. The pre-exponential factor $\tau_0^{-1}$ (sometimes called 
attempt frequency) was found to lie between $10^3$ and $10^6$ s$^{-1}$ whereas 
expected values are between $10^9$ and $10^{12}$ s$^{-1}$. Why one must apply 
fields of about 2 Teslas in order to measure two--level fluctuations which should be 
expected near zero field? What is the influence of the measurement technique on the 
sample? 

By measuring the electrical resistance of isolated Ni wires with diameters between 20 and 
40 nm, Hong and Giordano studied the motion of magnetic domain walls \cite{Hong95}. 
Because of surface roughness and oxidation, the domain walls of a single wire are trapped 
at pinning centres. The pinning barrier decreases with an increase in the magnetic field. 
When the barrier is sufficiently small, thermally activated escape of the wall occurs. This 
is a stochastic process which can be characterised by a switching (depinning) field 
distribution. A flattening of the temperature dependence of the mean switching field and a 
saturation of the width of the switching field distribution (rms. deviation $\sigma$) were observed 
below about 5 K. The authors proposed that a domain wall escapes from its pinning sites 
by thermal activation at high temperatures and by quantum tunnelling 
below $T_{\rm c} \sim$ 5 K.

These measurements pose several questions: What is the origin of the pinning center 
which may be related to surface roughness, impurities, oxidation etc.? The sweeping rate 
dependence of the depinning field, as well as the depinning probability, could not be 
measured even in the thermally activated regime. Therefore, it was not possible to check 
the validity of the Ne\'el--Brown model \cite{N_B} or to compare measured and predicted rms. deviations $\sigma$. Finally, a crossover temperature $T_{\rm c}$ of about 5 K is three orders of magnitude 
higher than $T_{\rm c}$ predicted by current theories.

Later, Wernsdorfer {\it et al.} published results obtained on nanoparticles synthesised by arc discharge, with dimensions between 10 and 30 nm \cite{Wernsdorfer97a}. These particles 
were single crystalline, and the surface roughness was about two atomic layers. Their 
measurements showed for the first time that the magnetisation reversal of a ferromagnetic 
nanoparticle of good quality can be described by thermal activation over a single--energy 
barrier as proposed by Ne\'el and Brown \cite{N_B}. The activation volume, which is the volume of 
magnetisation overcoming the barrier, was very close to the particle volume, predicted for 
magnetisation reversal by uniform rotation. No quantum effects were found down to 0.2 
K. This was not surprising because the predicted crossover temperature is $T_{\rm c} \sim$ 20 
mK. The results of Wernsdorfer {\it et al.} constitute the preconditions for the experimental 
observation of MQT of magnetisation on a single particle.

Just as the results obtained with Co nanoparticles \cite{Wernsdorfer97a}, a quantitative 
agreement with the Ne\'el--Brown model of magnetisation reversal was found on BaFeO 
nanoparticles \cite{Wernsdorfer97d}. Although, strong deviations from this model were 
evidenced for the smallest particles containing about $10^5 \mu_{\rm B}$ and for temperatures 
below 0.4 K. These deviations are in good agreement with the theory of macroscopic 
quantum tunnelling of magnetisation. The main results are reviewed in the following 
section.

Other low temperature techniques which are adapted 
to single particle measurements are Hall probe magnetometry~\cite{Geim98}, 
magnetometry based on the giant magnetoresistance or spin-dependent 
tunneling with Coulomb blockade~\cite{Gueron99}.

\section{Example: Magnetisation reversal by thermal activation and quantum 
tunnelling in BaFeCoTiO nanoparticles}

This section presents a brief discussion of individual particle measurements suggesting 
quantum effects at low temperature. 
In order to confirm the single--domain character of the particles, the magnetisation reversal 
as a function of the applied field direction were studied for each particle. For Co and 
BaFeCoTiO nanoparticles with diameters between 10 and 30 nm, it was found that the 
angular dependence of the switching field $H_{sw}^0$ agrees well with the model of 
Stoner and Wohlfarth \cite{St-W48} taking into account mainly second and small fourth order 
anisotropy terms. The effective magnetocrystalline anisotropy field (the main 
second order anisotropy term), found by these measurements for particles with a Co$_{0.8}$Ti$_{0.8}$ 
substitution, is about 0.4 T. A detailed discussion of the angular dependence of 
$H_{\rm sw}^0$ is presented in Ref. \cite{Bonet 1999}.

The influence of temperature and time on the statistics of the magnetisation reversal was 
studied by waiting time and switching field experiments. Both types of measurements 
were studied as a function of the applied field direction. Except in some special cases at 
very low temperatures, these measurements were in complete agreement with the 
Ne\'el--Brown theory \cite{N_B}: (i) exponential probabilities of not-switching 
with mean waiting times 
following an Arrhenius law; (ii) mean switching fields and widths of the switching field 
distribution following the model of Kurkij$\ddot{a}$rvi \cite{Kurkijarvi72}.

This agreement could be confirmed by studying the angular dependence of the anisotropy 
barrier $E_0(\theta)$ following roughly the prediction of the Stoner and Wohlfarth \cite{St-W48}. The 
number of spins $S$ in the nanoparticle can be 
estimated by $S \sim E_0 / (2 \mu_{\rm B} 
\mu_0 H_{\rm sw}^0)$. We found values of $S \sim 10^6$ to $10^5$ depending on the 
particles size. Finally, the angular dependence of $\tau_0$ followed well the prediction of 
Coffey {\it et al.} \cite{Coffey}. The phenomenological damping constant $\alpha$ from Gilbert's 
equation could be found: $\alpha \sim 10^{-1}$ in the case of Co and 
$\alpha \sim 10^{-2}$ in the case of BaFeCoTiO \cite{Coffey98}. 
Such values are expected for metallic and 
insulating particles, respectively.

Below 0.4 K, several of the smallest particles showed strong deviations from the Ne\'el--
Brown model. These deviations were a saturation of the thermal dependence of 
$H_{\rm sw}$ and $\sigma$, and a faster field sweeping rate dependence of $H_{\rm sw}$ 
than given by the Ne\'el--Brown model. In order to investigate the possibility that these low 
temperature deviations are due to an escape from the metastable potential by MQT, a 
common method is to replace the real temperature $T$ by an effective temperature 
$T^{\ast}(T)$ in order to restore the scaling plot. In the case of MQT, $T^{\ast}(T)$ should 
saturate at low temperatures. Indeed, the ansatz of $T^{\ast}(T)$, can restore unequivocally the 
scaling plot demonstrated by a straight master curve. The flattening of $T^{\ast}$ corresponds 
to a saturation of the escape rate $\Gamma$ which is a necessary signature of MQT. As 
measurements at zero temperature are impossible, the effective temperature at the lowest 
measuring temperature can be investigated and one can define the crossover temperature 
between thermally activated and the quantum regime by $T_{\rm c} = T^{\ast}$ at lowest measuring 
temperature.

The measured angular dependence of $T_{\rm c}(\theta)$ is in excellent agreement with the 
prediction given by equation (1) (Fig. 2). The normalisation value $T_{\rm c}(45^{\circ})$ = 0.31 K compares well with the theoretical value of about 0.2 K. This quantitative 
agreement of the crossover temperature versus an external parameter suggest MQT in 
these BaFeCoTiO particles. 

Measurements on other particles with anisotropies 
between $H_{\rm a}$ = 0.3 T and 1 T 
showed that $T_{\rm c}(\theta)$ is about proportional 
to $H_{\rm a}$. Furthermore, for a given 
anisotropy field $H_{\rm a}$, $T_{\rm c}(\theta)$ decreased for bigger particles which is also in 
agreement with equation (1) as $T_{\rm c}(\theta)$ is proportion to $\varepsilon^{1/4}$ 
(for a larger particle, $\varepsilon$ must be smaller in order to measure a magnetisation switching in 
the same time window, which is fixed by the field sweeping rate. Finally, 
the width of the switching field distribution should be constant in the quantum regime 
which was also verified.

The test of the validity for $\theta\rightarrow 0^{\circ}$ and $\theta\rightarrow 
90^{\circ}$ would be interesting. 
However, the limit $\theta\rightarrow 90^{\circ}$ is 
experimentally very difficult as the measurable signal at the magnetisation reversal 
becomes smaller and smaller for $\theta\rightarrow 90^{\circ}$. Furthermore, by using 
the micro--SQUID technique, one can only sweep the applied magnetic field in the plane of 
the SQUID, i.e. in order to measure in the limit of $\theta\rightarrow 0^{\circ}$, the easy 
axis of magnetisation needs to be in the plane of the SQUID which has not yet been 
achieved for a small particle. 

\begin{figure}[t]
\begin{center}\leavevmode
\includegraphics[width=0.8\linewidth]{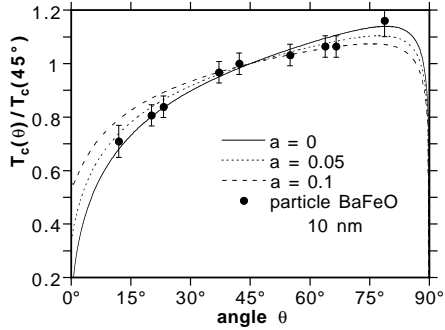}
\caption{Angular dependence of the crossover temperature $T_{\rm c}$ for BaFeCoTiO particle. The lines are given by equation (1) for different values of the ratio $a = H_{\perp}/H_{\parallel}$ The experimental data are normalised by $T_{\rm c}(45^{\circ})$ = 0.31 K..}
\label{fig2}
\end{center}
\end{figure}

In the case of 20 nm Co particles, no clear quantum effects were found down to 0.2 K. 
This was not surprising for two reasons: (i) the calculated crossover temperature for such 
a particle is smaller than 0.2 K and (ii) the dissipation effects due to conduction electrons 
may strongly reduce quantum effects which should not be the case in insulating 
BaFeCoTiO particles having a very small damping factor \cite{Coffey98}.

Although the above measurements are in good agreement with MQT theory, we should 
not forget that the MQT is based on several strong assumptions. Among them, there is the 
assumption of a giant spin, i.e. all magnetic moments in the particle are rigidly coupled 
together by strong exchange interaction. This approximation might be good in the 
temperature range where thermal activation is dominant but is it not yet clear if this 
approximation can be made for very low energy barriers. Future measurements will tell us 
the answer.

The proof for MQT in a magnetic nanoparticle could be the observation of level 
quantisation of its collective spin state which was recently evidenced in molecular 
Mn$_{12}$ and Fe$_8$ clusters having a collective spin state $S = 10$. Also the 
quantum spin phase or Berry phase associated with the magnetic spin $S = 10$ of a 
Fe$_8$ molecular cluster was evidenced \cite{Wernsdorfer99}. In the case of BaFeCoTiO 
particles with S = $10^5$, the field separation associated with the level quantisation is 
rather small: $\Delta H = H_{\rm a}/ 2 S \sim$ 0.002 mT. Future measurements should focus 
on the level quantisation of collective spin states of $S = 10^2$ to $10^4$ and their 
quantum spin phases

%

\end{document}